\documentclass[conference]{IEEEtran}
\IEEEoverridecommandlockouts
\usepackage{amssymb,amsmath}
\usepackage{graphics}
\usepackage{float}
\usepackage{graphicx}
\usepackage{amsmath}
\usepackage{cite}
\usepackage{color}
\usepackage{dsfont}
\usepackage{bm}
\usepackage{upgreek}
\usepackage{arydshln}
\usepackage{enumerate}
\usepackage{amsthm}
\usepackage{graphicx}
\usepackage{threeparttable}
\usepackage{caption}
\usepackage{multirow}
\usepackage{color}
\usepackage{xfrac}
\usepackage{array}
\usepackage{nomencl}
\usepackage{hyphenat}
\usepackage{cite}
\usepackage{amsmath,amssymb,amsfonts}
\usepackage{algorithmic}
\usepackage{graphicx}
\usepackage{textcomp}
\usepackage{xcolor}

\newcommand{\be}{\begin{equation}}
\newcommand{\ee}{\end{equation}}

\newcommand{\norm}[1]{ || #1 ||}
\newcommand{\mb}[1]{\mathbf{#1}}
\newcommand{\bs}[1]{\boldsymbol{#1}}
\newcommand{\lbr}{\left\lbrace}

\newcommand{\virg}[1]{\textquotedblleft#1\textquotedblright}

\newcommand{\vecl}[1]{\mathrm{vec}_l(#1)}

\newcommand{\tonde}[1]{\left( #1 \right)  }
\newcommand{\quadre}[1]{\left[  #1 \right]  }

\begin{document}

\title{Semiparametric Stochastic CRB for DOA Estimation in Elliptical Data Model\\
\thanks{The work of Stefano Fortunati has been partially supported by the Air Force Office of Scientific Research under award number FA9550-17-1-0065.}
}

\author{\IEEEauthorblockN{Stefano Fortunati, Fulvio Gini, Maria S.\ Greco}
	\IEEEauthorblockA{Dept. of Information Engineering, University of Pisa\\
		\{s.fortunati,f.gini,m.greco\}@iet.unipi.it}}

\maketitle

\begin{abstract}
This paper aims at presenting a numerical investigation of the statistical efficiency of the MUSIC (with different covariance matrix estimates) and the IAA-APES Direction of Arrivals (DOAs) estimation algorithms under a general Complex Elliptically Symmetric (CES) distributed measurement model. Specifically, the density generator of the CES-distributed data snapshots is considered as an additional, infinite-dimensional, nuisance parameter. To assess the efficiency in the considered semiparametric setting, the Semiparametric Stochastic Cram\'{e}r-Rao Bound (SSCRB) is adopted as lower bound for the Mean Square Error (MSE) of the DOA estimators.    
\end{abstract}

\begin{IEEEkeywords}
DOA estimation, Semiparametric model, Semiparametric Stichastic Cram\'{e}r-Rao Bound.
\end{IEEEkeywords}

\section{Introduction}
Estimation of the Direction of Arrival (DOA) of a certain number of sources using an array of active or passive sensors is a standard signal processing problem. There is a huge literature on this topic covering different aspects: parametric and non-parametric estimation methods, deterministic and random signal models and so on (see e.g. \cite{krim,vantrees} and the references therein). Along with the statistical signal models and the related estimation methods, considerable efforts have been dedicated to the derivation of suitable performance bounds for DOA estimation (see e.g. the standard references \cite{Stoica_CRB_2,Stoica_CRB} and the overview provided in \cite{Delmas}). However, most of the works on array processing and, in particular, the ones on lower bounds, assume a Gaussian model for the collected data snapshots. A valuable exception is represented by the paper \cite{Esa_DOA}, where the signal model is statistically characterized using the set of the Complex Elliptically Symmetric (CES) distributions \cite{Esa}. The CES class is a wide family of non-Gaussian distributions that encompasses the Gaussian, the Generalized Gaussian, the $t$-distribution and all the Compound Gaussian distributions as special cases. As detailed in \cite{Esa} and \cite{Sang}, many experimental evidences have shown their ability to characterize the heavy-tailed behaviour of real datasets collected in different applications such as radar/sonar, indoor/outdoor wireless communications or seismic data processing. Note that, in all these examples, the DOA estimation is a key aspect.   

The main goal of this paper is then to provide a numerical investigation of the \textit{statistical efficiency} of some of the most widely used DOA estimation algorithms by dropping the classical Gaussianity assumption in favour of a more general model in which \textit{i}) the collected snapshots $\{\mb{z}\}_{l=1}^L$ are assumed to be CES-distributed with \textit{unknown} density generator and \textit{ii}) the signal and disturbance components are uncorrelated. To handle the additional unknown \textit{infinite-dimensional} parameter, that is the desity generator characterizing the actual CES data distribution, we exploit our recent findings on the Semiparametric Stochastic Cram\'{e}r-Rao Bound (SSCRB) \cite{For_Com_SCRB}. 

To make the paper as self-contained as possible, Section II provides a very short introduction on CES distribution. The assumed measurement model and the related SSCRB on the estimation of the DOAs of $K$ narrowband sources are introduced in Section III. Sections IV and V discuss the DOA estimation algorithms while their efficiency with respect to (w.r.t.) the SSRCB is investigated in Section VI. Some concluding remark is collected in Section VII.

\section{Preliminaries: essentials on CES distributions}
\label{Sec_CES}
This section presents some basic properties of CES distributed random vectors that will be used in the remaining of the paper. For a complete and insightful discussion on CES distributions, we refer the reader to \cite{Esa} and references therein.

An $N$-dimensional CES-distributed random vector $\mb{z}$ is fully characterized by its mean vector $\bs{\mu} \in \mathbb{C}^N$, its scatter matrix $\bs{\Sigma} \in \mathbb{C}^{N \times N}$ and its density generator $h \in \mathcal{G}$, where $\mathcal{G}$ is a suitable set of functions. Under the \textit{absolutely continuous} assumption, i.e. when the scatter matrix has full rank, the pdf of a CES-distributed vector $\mb{z}\sim CES_N(\mb{z};\bs{\mu},\bs{\Sigma},h)$ is:
\be
\label{true_CES}
\begin{split}
	p_{Z}(\mb{z}|\bs{\mu},\bs{\Sigma}, h) 
	=|\bs{\Sigma}|^{-1} h \left(  (\mb{z}-\bs{\mu})^{H}
	\bs{\Sigma}^{-1}(\mb{z}-\bs{\mu}) \right).
\end{split}
\ee
Moreover, $\mb{z}$ satisfies the circularity property, i.e. $(\mb{z}-\bs{\mu}) =_d e^{j\vartheta}(\mb{z}-\bs{\mu}), \; \forall \vartheta \in \mathbb{R}$. Any CES-distributed vector $\mb{z}$ admits the following representation:
\be
\label{CSRT_dec}
\mb{z} =_d \bs{\mu} + \sqrt{\mathcal{Q}}\bs{\Sigma}^{1/2}\mb{u},
\ee
where $\mb{u} \sim \mathcal{U}(\mathbb{C}S^N)$ is a complex random vector uniformly distributed on the unit complex $N$-sphere $\mathbb{C}S^N$ and $\mathcal{Q}$ is the so-called \textit{2nd-order modular variate}, such that (s.t.):
\be
\label{somv}
\mathcal{Q} =_d Q \triangleq (\mb{z}-\bs{\mu})^H \bs{\Sigma}^{-1}(\mb{z}-\bs{\mu}),
\ee
whose pdf is given by:
\be
\label{CSS_Q_pdf}
p_{\mathcal{Q}}(q) = \pi^{N}\Gamma(N)^{-1}q^{N-1} h (q).
\ee
From \eqref{CSRT_dec} and by exploiting the properties of $\mb{u}$ \cite[Lemma 1]{Esa}, we have that the covariance matrix of the CES-distributed vector $\mb{z}$ is $\mb{M} \triangleq E\{(\mb{z}-\bs{\mu})(\mb{z}-\bs{\mu})^H\} = N^{-1}E\{\mathcal{Q}\}\mb{\Sigma}$.    

In order to remove the well-known scale ambiguity, we impose a constraint on the functional form of the density generator $h$. Following the same procedure adopted in \cite{miss_sb,For_Com_SCRB}, we assume that $h \in \mathcal{G}$ is parameterized in order to satisfy the constraint:
\be
\label{const}
E\{\mathcal{Q}\}=\pi^{N}\Gamma(N)^{-1}\int_{0}^{+\infty}q^{N-1} h (q)dq=N.
\ee
As a consequence of \eqref{const}, the scatter matrix $\bs{\Sigma}$ equates the covariance matrix $\mb{M}$ of $\mb{z}$ \cite[Sec. III.C]{Esa}. For further reference, we define the set $\bar{\mathcal{G}} \subset \mathcal{G}$ as the set of all the density generators satisfying the constraint in \eqref{const}. Moreover, all the expectation operator w.r.t. the \virg{constrained} pdf of the second-order modular variate in \eqref{somv} will be indicated as $\bar{E}\{\cdot\}$.

\section{The measurement model}
Suppose to have a uniformly linear array (ULA) of $N$ omnidirectional sensors and $K$ narrowband sources characterized by $K$ spatial frequencies $\{\nu_k\}_{k=1}^K$. For the ULA configuration, the spatial frequency $\nu_k$ and the (conic) angle of arrival $\gamma_k$ is linked by $\nu_k = d/\lambda\sin(\gamma_k)$ where $d$ is the spacing between the sensor and $\lambda$ is the wavelength of the transmitted signal. The adoption of $\nu_k$, instead of $\gamma_k$, as direction parameter allows us to discard the non-linearity due to the $\sin$ function and the dependence on the \virg{system-dependent} parameters $d$ and $\lambda$. For a ULA, the steering vector can be expressed as $\mb{a}(\nu_k)=(1,e^{j2\pi\nu_k},\ldots,e^{j2\pi(N-1)\nu_k})^T$.
Furthermore, by defining $\bs{\nu}=(\nu_1,\ldots,\nu_K)$ as the vector collecting all the source spatial frequencies, we can define the steering matrix $\mb{A}(\bs{\nu})\triangleq [\mb{a}(\nu_1)|\cdots|\mb{a}(\nu_K)] \in \mathbb{C}^{N \times K}$ as the matrix whose $k$-th column is the steering vector related to the $k$-th spatial frequency.
In the rest of the paper, we assume to have a set of $L$ zero-mean, independent and identically CES-distributed (i.i.d.) data snapshots $\{\mb{z}_l\}_{l=1}^L$, s.\ t.:
\be
\label{snap_dist}
\mathbb{C}^{N} \ni \mb{z}_l \sim CES_N(\mb{z}_l;\mb{0},\bs{\Sigma}(\bs{\theta}_0),h_0),
\ee
where $h_0$ is the \textit{true}\footnote{Note that we use the subscript $0$ to distinguish between the \textit{true} density generator $h_0 \in \bar{\mathcal{G}}$ and a \textit{generic} function $h$ in $\bar{\mathcal{G}}$. This notation will be adopted for any other vector, matrix or function in the paper.}, but generally \textit{unknown}, density generator. The scatter matrix is:
\be
\mathbb{C}^{N \times N} \ni \bs{\Sigma}_0 \equiv \bs{\Sigma}(\bs{\theta}_0) = \mb{A}_0\bs{\Gamma}_0\mb{A}_0^H + \sigma^2_0\mb{I}_N,
\ee
where $\mb{A}_0 \equiv \mb{A}(\bs{\nu}_0)$, $\bs{\Gamma}_0 \in \mathbb{C}^{K \times K}$ is the source covariance matrix, $\mb{I}_N$ is the identity matrix of dimension $N \times N$ and $\sigma^2_0$ is the \virg{noise} power. The \textit{true}, but again generally \textit{unknown}, parameter vector $\bs{\theta}_0$, is defined as:
\be
\label{theta}
\bs{\theta}_0 \triangleq [\bs{\nu}_0^T,\bs{\zeta}_0^T,\sigma_0^2]^T \in \mathbb{R}^{K + N^2 +1},
\ee
and the vector $\bs{\zeta}_0$ is the $N^2$-dimensional real vector such that:
\be
\bs{\zeta}_0 \triangleq \quadre{\mathrm{diag}(\bs{\Gamma}_0)^T,\mathrm{vec}_l(\mathrm{Re}(\bs{\Gamma}_0))^T,\mathrm{vec}_l(\mathrm{Im}(\bs{\Gamma}_0))^T}^T,
\ee
where the operator $\vecl{\cdot}$ selects all the entries strictly below the main diagonal of $\bs{\Gamma}_0$ taken in the same column-wise order as the ordinary $\mathrm{vec}(\cdot)$ operator \cite[Sec. 2.4]{Complex_M} while $\mathrm{diag}(\bs{\Gamma}_0)$ is a column vector collecting the diagonal elements of $\bs{\Gamma}_0$.
\textit{Remark 1}: It is immediate to verify that, when the density generator is $h_0 = \exp(-t)$, i.e. when the snapshot $\mb{z}_l$ is Gaussian-distributed, the signal model in \eqref{snap_dist} is the classical random signal model used e.g. in \cite{Stoica_CRB_2} and \cite{Stoica_CRB}:
\be
\mb{z}_l = \mb{A}(\bs{\nu}_0) \mb{s}_l + \mb{w}_l,
\ee
where $\mb{s}_l \in \mathbb{C}^K$ is the, zero mean, circular Gaussian signal random vector whose covariance matrix is $\bs{\Gamma}_0 = E\{\mb{s}_l\mb{s}_l^H\}$ and $\mb{w}_l \sim CN(\mb{0},\sigma_0^2\mb{I}_N)$ is the white Gaussian measurement noise. This observation motivates the characterization of the parameter $\sigma_0^2$ in the general model \eqref{snap_dist} as noise power.
\subsection{The Stochastic CRB}
As largely discussed in the array processing literature, we are generally interested in the estimation of $\bs{\nu}_0$, that is the vector of the spatial frequencies, while the other two terms in the parameter vector $\bs{\theta}_0$ in \eqref{theta}, i.e.\ the signal covariance $\bs{\zeta}_0$ and the noise power $\sigma_0^2$ have to be considered as \textit{nuisance} parameters. A CRB for the estimation of $\bs{\nu}_0$ in the presence of the (finite-dimensional) nuisance parameter vectors $\bs{\zeta}_0$ and $\sigma_0^2$ has been discussed in \cite{Stoica_CRB} under the Gaussian model assumption discussed in Remark 1. This bound, called the Stochastic CRB (SCRB), is given by:
\be
\label{SCRB}
\mathrm{SCRB}(\bs{\nu}_0|\bs{\zeta}_0,\sigma_0^2) = \frac{\sigma_0^2}{2L}\mb{C}(\bs{\nu}_0,\bs{\zeta}_0)^{-1},
\ee
\be
\label{C_mat}
\mb{C}(\bs{\nu}_0,\bs{\zeta}_0) \triangleq \mathrm{Re}\tonde{\mb{D}_0^H \Pi^{\perp}_{\mb{A}_0} \mb{D}_0}\odot \tonde{\bs{\Gamma}_0\mb{A}_0^H\mb{\Sigma}_0^{-1}\mb{A}_0\mb{\Gamma}_0}^T,
\ee
where $\odot$ is the Hadamard product, $\mb{D}_0 \triangleq \quadre{\mb{d}_{0,1},\cdots,\mb{d}_{0,K}}$ where $\mb{d}_{0,k}\triangleq \left. d\mb{a}(\nu_k)/d\nu_k \right|_{\nu_k = \nu_{0,k}}$ and 
\be
\Pi^{\perp}_{\mb{A}_0} = \mb{I}_N - \mb{A}_0(\mb{A}^H_0\mb{A}_0)^{-1}\mb{A}_0^H.
\ee
\subsection{The Semiparametric Stochastic CRB}
Following the theoretical results obtained in our recent work \cite{For_Com_SCRB}, we can take these two steps further:
\begin{enumerate}
 \item We drop the Gaussianity assumption in favour of the more general CES model in \eqref{snap_dist},
 \item By relying on the \textit{semiparametric} framework\footnote{The reader that is not familiar with the semiparametric theory may have a look at the books \cite{BKRW} and \cite{Tsiatis} and to the wide statistical literature available on this topic. Moreover, we may suggest the reader to look into our recent works \cite{For_EUSIPCO,For_SCRB,For_Com_SCRB} where the semiparmaetric nature of the CES distributions has been analysed.}, we consider as additional \textit{nuisace} parameter the density generator $h_0$ itself. 
\end{enumerate}
Roughly speaking, the semiparametric framework addressed in \cite{For_EUSIPCO,For_SCRB,For_Com_SCRB} allow us to derive a lower bound on the performance of any estimator of the vector of the spatial frequencies $\bs{\nu}_0$ when the signal covariance $\bs{\zeta}_0$, the noise power $\sigma_0^2$ and even the density generator $h_0$ are unknown. The only assumption used to derive this bound, that we call Semiparametric SCRB (SSCRB), is that the data snapshots are CES-distributed as in \eqref{snap_dist}. As proved in \cite{For_Com_SCRB}, the SSCRB can be expressed as:
\be
\label{SSCRB}
\mathrm{SSCRB}(\bs{\nu}_0|\bs{\zeta}_0,\sigma_0^2, h_0) = \frac{N(N+1)\sigma_0^2}{2L\bar{E}\{\mathcal{Q}^2\psi_0(\mathcal{Q})^2\}} \mb{C}(\bs{\nu}_0,\bs{\zeta}_0)^{-1},
\ee
where the matrix $\mb{C}(\bs{\nu}_0,\bs{\zeta}_0)$ is the one given in \eqref{C_mat}, $\mathcal{Q}$ is the 2nd-order modular variate defined in \eqref{somv} and $\psi_0(t) \triangleq d \ln h_0(t)/dt$ where $h_0$ is the actual density generator of the data snapshots.
 
\section{MUSIC with robust scatter matrix estimators}
After having introduced the measurement model and the related SSCRB, we now present the MUSIC estimation algorithm whose efficiency, w.r.t. the SSCRB, will be investigated by simulations in Section VI. The MUSIC estimator of the vector $\bs{\nu}_0$ of spatial frequencies is given by (see e.g. \cite{MUSIC}):
\be
\label{Mus_al}
\hat{\bs{\nu}} = \underset{\nu}{\mathrm{argmax}} \quadre{\sum\nolimits_{n=K+1}^N |\mb{a}(\nu)^H\hat{\mb{v}}_n|^2}^{-1},
\ee  
where $\mb{a}(\nu)$ is the steering vector and $\{\hat{\mb{v}}_n\}_{n=K+1}^N$ are the $N-K$ eigenvectors corresponding to the $N-K$ smallest eigenvalues of the estimated data covariance matrix $\hat{\bs{\Sigma}}$. It is worth underling that the MUSIC algorithm does not require the a-priori knowledge of the functional form of the actual density generator $h_0$, that is generally unknown, so it can be applied in the considered semiparametric framework. Let us now focus our attention on the estimation of the covariance matrix $\hat{\bs{\Sigma}}$. Also in this case, we have to rely on estimators that do not make use of a-priori information on $h_0$. As a consequence, Maximum Likelihood estimator is not an option. Here, we list five \virg{semiparametric} estimators that we are going to take into account in the efficiency study of the MUSIC algorithm.
\subsection{The Sample Covariance Matrix (SCM)}
The well-known SCM estimate of $\bs{\Sigma}$ is given by: 
\be
\label{SCM}
\hat{\bs{\Sigma}}_{SCM} \triangleq \frac{1}{L} \sum\nolimits_{l=1}^{L} \mb{z}_l\mb{z}_l^H.
\ee 
$\hat{\bs{\Sigma}}_{SCM}$ is the ML estimator when the data are Gaussian, but its performance drastically decreases in heavy-tailed scenarios.
\subsection{The Normalized (or Sign) and the Kendall's Tau SCM}
Let us define the \textit{spatial sign function} \cite{visuri} as:
\be
\mb{v}(\mb{z}) \triangleq \left\lbrace \begin{array}{cc}
	\mb{z}/\norm{\mb{z}},& \mb{z} \neq \mb{0}\\
	\mb{0},& \mb{z} = \mb{0}
\end{array} \right. .
\ee
Then, the Normalized SCM (NSCM) \cite{Gini1,NSCM} and the Kendall's Tau SCM \cite{visuri} are simply defined as:
\be
\label{NSCM}
\hat{\bs{\Sigma}}_{NSCM} \triangleq \frac{1}{L} \sum\nolimits_{l=1}^{L} \mb{v}(\mb{z}_l)\mb{v}(\mb{z}_l)^H,
\ee
\be
\label{TSCM}
\hat{\bs{\Sigma}}_{KT} \triangleq \frac{1}{L(L-1)} \sum_{i=1}^{L}\sum_{j=1}^{L} \mb{v}(\mb{z}_i-\mb{z}_j)\mb{v}(\mb{z}_i-\mb{z}_j)^H.
\ee
The use of these two nonparametric estimators in DOA estimation problem has been firstly discussed in \cite{visuri}.
\subsection{Tyler's and Huber's $M$-estimators}
The Tyler's and Huber's estimates are the convergence points of the following iterative algorithm: 
\be
\label{C_Tyler}
\hat{\bs{\Sigma}}^{(k+1)} = \frac{1}{L}\sum_{l=1}^{L} \varphi(\mb{z}_l^H(\hat{\bs{\Sigma}}^{(k)})^{-1}\mb{z}_l)\mb{z}_l\mb{z}_l^H,
\ee
where the starting point is $\hat{\bs{\Sigma}}^{(0)} = \mb{I}_N$. The weight function $\varphi(t)$ for Tyler's estimator is defined as (see e.g. \cite{Esa}):
\be
\label{w_Tyler}
\varphi_{Tyler}(t)=N/t,
\ee 
whereas the one for Huber's estimator is given by:
\be
\label{w_Huber}
\varphi_{Hub}(t)= \lbr
\begin{array}{cc}
	1/b & t\leqslant \delta^2 \\
	\delta^2/(tb) & t> \delta^2
\end{array} \right.,
\ee 
where $q = F_{\chi_N^2}(2\delta^2) \in (0,1]$ is a tuning parameter and $F_{\chi_N^2}(\cdot)$ indicates the distribution of a chi-squared random variable with $N$ degrees of freedom. The parameter $b$ is usually chosen as $b=F_{\chi_{N+2}^2}(2\delta^2)+\delta^2(1-b)/N$ \cite{Esa}. Note that Tyler's estimator is the minimax robust $M$-estimator of the scatter matrix for CES-distributed data, while Huber's one represents a compromise between the robustness of Tyler's estimator (that can be obtained for $q=0$) and the efficiency at Gaussianity of the SCM ($q=1$) \cite{Esa}.

\section{The IAA-APES algorithm}
The Iterative Adaptive Approach for Amplitude and Phase EStimation (IAA-APES) is a least squares-based algorithm that, as the MUSIC algorithm, does not exploit any information on the data distribution but, unlike MUSIC, does not rely of the estimation of the snapshot covariance matrix \cite{APES}. Here, a short description of the IAA-APES method is provided. For additional details and and discussions, we refer the reader to \cite{APES}. Let $\Omega = \{ \nu_g \}_{g=1}^G$ be a grid of possible spatial frequencies and let $\mb{P}$ be a diagonal matrix whose diagonal elements $\{ P_{gg} \}_{g=1}^G$ are the powers of the potential sources with spatial frequencies $\{ \nu_g \}_{g=1}^G$. Following \cite{APES}, we introduce the matrix 
\be
\label{Apes_Q}
\mb{Q}(\nu_g) \triangleq \mb{R} - P_{gg}\mb{a}(\nu_g)\mb{a}(\nu_g)^H,
\ee
where $\mb{R} \triangleq \mb{A}(\Omega)\mb{P}\mb{A}(\Omega)^H$ and $\mb{A}(\Omega)$ is a matrix whose columns are the steering vectors for each spatial frequency in the grid $\Omega$. The IAA-APES cost function is defined as:
\be
\sum\nolimits_{l=1}^{L}\norm{\mb{z}_l - s_{g,l}\mb{a}(\nu_g)}^2_{\mb{Q}(\nu_g)^{-1}},
\ee
and, by minimizing w.r.t. the signal parameter $s_{g,l}$, we get:
\be
\label{Apes_s}
\hat{s}_{g,l} = \frac{\mb{a}(\nu_g)^H\mb{Q}(\nu_g)^{-1}\mb{z}_l}{\mb{a}(\nu_g)^H\mb{Q}(\nu_g)^{-1}\mb{a}(\nu_g)}.
\ee 
Finally, the $K$ sources spatial frequencies can be identified as the $K$ elements of $\Omega = \{ \nu_g \}_{g=1}^G$ whose indices characterize the $K$ smallest diagonal elements of: 
\be
\hat{P}_{gg} \triangleq \frac{1}{L}\sum\nolimits_{l=1}^{L} |\hat{s}_{g,l}|^2.
\ee
Note that, since to implement \eqref{Apes_s}, we need an estimate of $\{ P_{gg} \}_{g=1}^G$ (see the definition of the matrix $\mb{Q}$ in \eqref{Apes_Q}), the estimation of $\hat{s}_{g,l}$ has to be implemented in an iterative algorithm as detailed in \cite{APES}.

\section{Numerical results}
This section is dedicated to the numerical assessment of the efficiency of MUSIC and IAA-APES algorithms w.r.t. the SSCRB in \eqref{SSCRB}. In the following simulations, we assume to have two sources at spatial frequencies $\nu_1 = -0.1$ and $\nu_2 = 0.3$. The noise power $\sigma_0^2 = 1$ while the source covariance matrix is:
\be
\bs{\Gamma}_0 = \tonde{ \begin{array}{cc}\sigma_1^2 & \rho \sigma_1 \sigma_2\\ \rho \sigma_1 \sigma_2& \sigma_2^2 \end{array} } ,
\ee
with $\rho = 0.3$. In Figures \ref{fig:Fig1} and \ref{fig:Fig2} where the efficiency is assessed as function of the Signal-to-Noise ratio (SNR), $\sigma_1^2$ and $\sigma_2^2$ are chosen as $\sigma_1^2 = \sigma_0^2 \cdot 10^{(\mathrm{SNR}/10)}$ and $\sigma_2^2 = \sigma_0^2 \cdot 10^{((\mathrm{SNR}-10)/10)}$, while in Figures \ref{fig:Fig3} and \ref{fig:Fig4}, where the efficiency is assessed as function of the non-Gaussianity of the collected data, $\sigma_1^2$ and $\sigma_2^2$ are chosen according to $\mathrm{SNR}_1 = 15$dB and $\mathrm{SNR}_2 = 10$dB.  
In all our simulations, the number of snapshots is $L=3N$, $N=8$ and the number of Monte Carlo runs in $10^5$. The tuning parameter for the Huber's estimator is $q=0.6$. The IAA-APES algorithm has been implemented with a maximum number of iterations equal to 30. As Mean Square Error indices, we use:
\be
\varepsilon_\alpha \triangleq E\{\norm{(\hat{\bs{\nu}}_\alpha - \bs{\nu}_0)(\hat{\bs{\nu}}_\alpha - \bs{\nu}_0)^T}_F\},
\ee
where $\norm{\cdot}_F$ is the Frobenius norm and $\alpha = \{SCM, NSCM, KT, Huber, Tyler, IAA-APES\}$. As lower bound, we plot the following index:
\be
\varepsilon_{SSCRB} = \norm{\mathrm{SSCRB}(\bs{\nu}_0|\bs{\zeta}_0,\sigma_0^2, h_0)}_F.
\ee
The efficiency study has been conducted for two CES distributions: the complex $t$- and the Generalized Gaussian (GG) distributions. A brief description of these two distributions along with the relevant calculation needed to obtain a closed form for SSCRB in \eqref{SSCRB} is given below. 
\subsection{$t$-distributed data}
The pdf related to the complex $t$-distribution is \cite{For_Com_SCRB}: 
\be
\label{dg_t_dist}
h_0(t) = \frac{\Gamma(\lambda+N)}{\pi^{N}\Gamma({\lambda} )}\left( \frac{\lambda}{\eta}\right) ^{\lambda}\left( \frac{\lambda}{\eta} + t \right)^{-(\lambda+N)}, 
\ee
and then $\psi_0(t) = -(\lambda + N) (\lambda/\eta + t) ^ {-1}.$ From \eqref{CSS_Q_pdf}, we have that:
\be
p_{\mathcal{Q}}(q) = \frac{\Gamma(\lambda+N)}{\Gamma(N)\Gamma({\lambda} )}\left( \frac{\lambda}{\eta}\right)^{\lambda}q^{N-1}\left( \frac{\lambda}{\eta} + q \right)^{-(\lambda+N)}.
\ee
As discussed in Sec. \ref{Sec_CES}, we have to constrain the density generator of the $t$-distribution to satisfy the constraint in \eqref{const}. It is immediate to verify that the constraint is satisfied by choosing $\eta = \lambda/(\lambda-1)$. Note that, for small values of the shape parameter $\lambda \in (1,\infty)$ the $t$-distribution have tails heavier that the Normal one, while $\lambda \rightarrow \infty$ the $t$-distributed data tends to be Gaussian.  
Using the integral in \cite[pp. 315, n. 3.194 (3)]{Integrals}, we get:
\be
\label{exp_2}
\bar{E}\{\mathcal{Q}^2\psi(\mathcal{Q})^2\} =  \frac{ N (N+1) (\lambda + N)}{(N +\lambda +1)}.
\ee

\subsection{GG-distributed data}
The pdf related to the GG distribution is \cite[Sec. IV.B]{Esa}: 
\be
\label{dg_GG}
h_0(t) = \frac{s\Gamma(N)b^{-N/s}}{\pi^N\Gamma(N/s)}\exp\tonde{-\frac{t^2}{b}},
% \frac{\Gamma(\lambda+N)}{\pi^{N}\Gamma({\lambda} )}\left( \frac{\lambda}{\eta}\right) ^{\lambda}\left( \frac{\lambda}{\eta} + t \right)^{-(\lambda+N)} 
\ee
and then $\psi_0(t) = -sb^{-1}t^{s-1}.$ From \eqref{CSS_Q_pdf}, we have that:
\be
p_{\mathcal{Q}}(q) = \frac{sb^{-N/s}}{\Gamma(N/s)}q^{N-1}\exp\tonde{-\frac{q^2}{b}}.
\ee
The GG distribution could have heavier tails ($s<1$) and lighter tails ($s>1$) as compared to the Normal one ($s=1$). As discussed in \cite[Sec. IV.B]{Esa}, in order to satisfy the constraint is \eqref{const}, the scale parameter $b$ as to be chosen as $b=\quadre{N\Gamma(N/s)/\Gamma((N+1)/s)}^s$. Using the integral in \cite[pp. 370, n. 3.478 (1)]{Integrals}, we get:
\be
\label{exp_3}
\bar{E}\{\mathcal{Q}^2\psi(\mathcal{Q})^2\} =  N(N+s).
\ee

\begin{figure}[H]
	\centering
	\includegraphics[height=5cm]{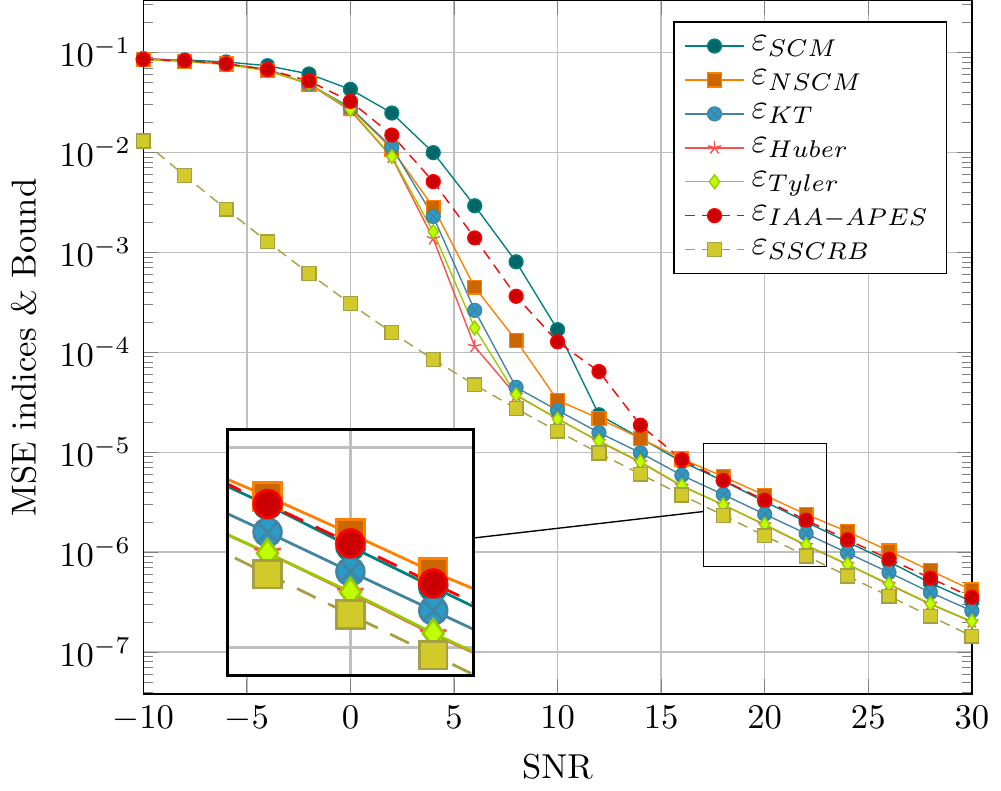}
	\caption{MSE and SSCRB vs SNR for \textit{t}-distributed data ($\lambda = 2$).}
	\label{fig:Fig1}
\end{figure}

\begin{figure}[H]
	\centering
	\includegraphics[height=5cm]{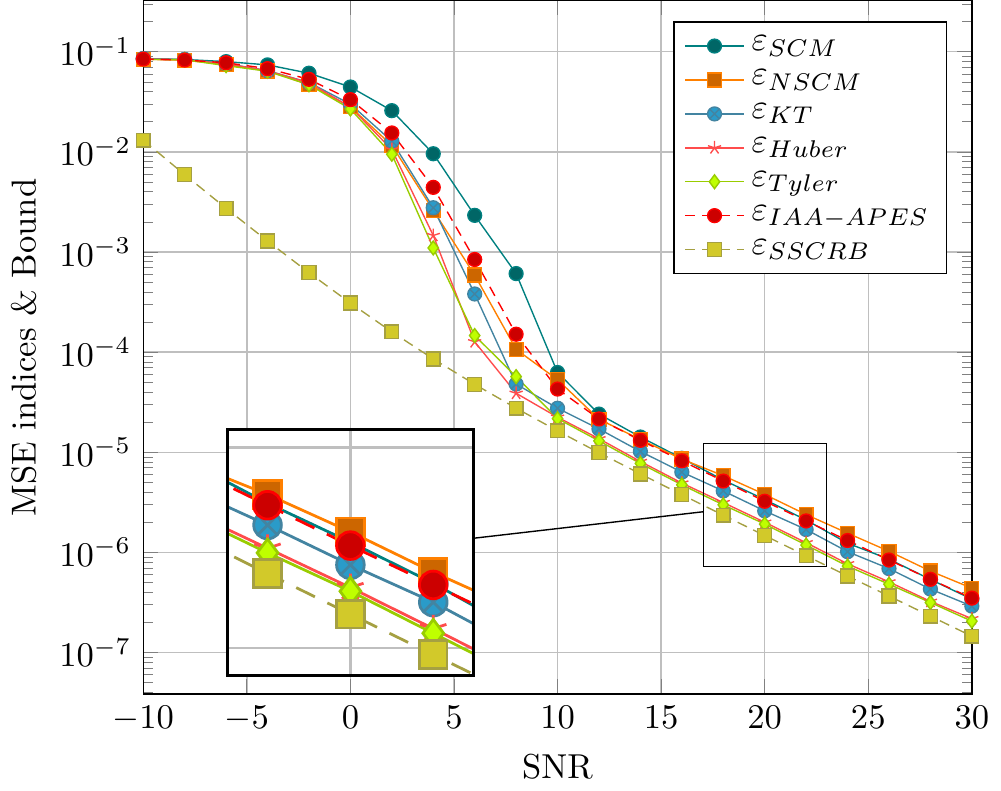}
	\caption{MSE and SSCRB vs SNR for GG data ($s=0.1$).}
	\label{fig:Fig2}
\end{figure}

\begin{figure}[H]
	\centering
	\includegraphics[height=5cm]{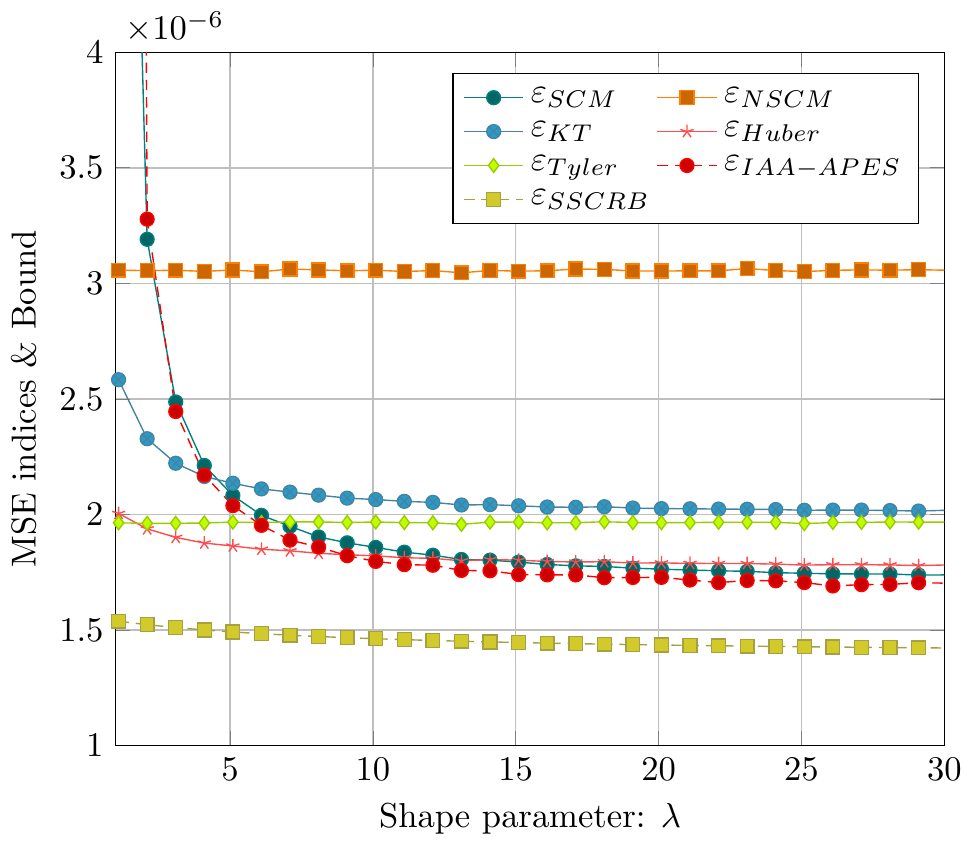}
	\caption{MSE and SSCRB vs $\lambda$ for \textit{t}-distributed data.}
	\label{fig:Fig3}
\end{figure}

\begin{figure}[H]
	\centering
	\includegraphics[height=5cm]{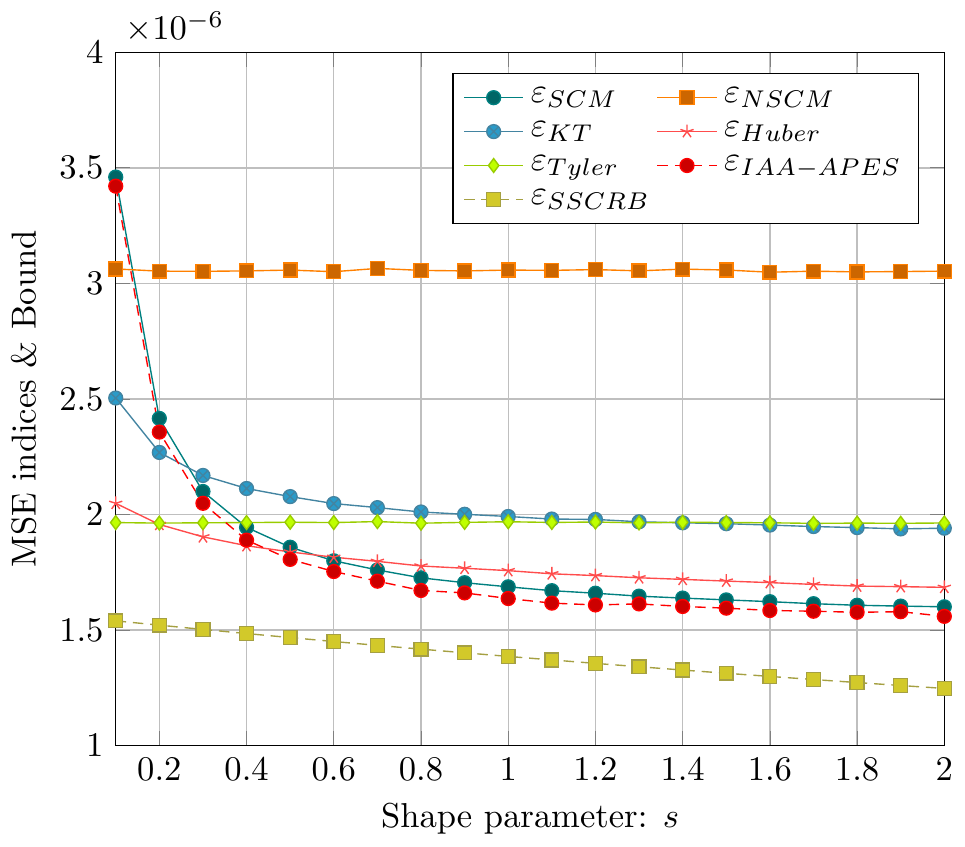}
	\caption{MSE and SSCRB vs $s$ for GG data.}
	\label{fig:Fig4}
\end{figure}

In Figures \ref{fig:Fig1} and \ref{fig:Fig2} the MSE of the MUSIC and IAA-APES and the related SSCRB are reported as function of the SNR, while in Figures \ref{fig:Fig3} and \ref{fig:Fig4} the efficiency w.r.t the SSCRB of the two DOA estimation algorithms is investigated as function of the non-Gaussianity of the collected data. Here, some observations:
\begin{itemize}
\item In the presence of heavy tailed data, the MUSIC-Tyler and the MUSIC-Huber algorithms present the best DOA estimation performance in both low and high SNR regimes. On the other hand, as expected, the MUSIC algorithm with the SCM performs poorly in non-Gaussian scenarios.
\item The performance of the IAA-APES algorithm are generally close to the one of the MUSIC-SCM algorithm and, in particular, it rapidly decreases as the spikiness of the data increases (see Figs. \ref{fig:Fig3} and \ref{fig:Fig4}).
\item Kendall's Tau SCM outperforms the NSCM on both the considered $t$- and GG-distributed data (Figs. \ref{fig:Fig3} and \ref{fig:Fig4}).
\item None of the considered DOA estimation algorithms is efficient w.r.t. the SSCRB. However, for a sufficiently large SNR value, MUSIC-Tyler and MUSIC-Huber algorithms are almost efficient. 	
\end{itemize} 
\vspace{-0.5cm}
\vspace{-0.5cm}
\vspace{-0.5cm}

\section{Concluding remarks}
In this paper, the statistical efficiency of the MUSIC (with different scatter matrix estimates) and the IAA-APES estimators has been assessed in the presence of CES-distributed data whose density generator is unknown and has been considered as an infinite-dimensional, nuisance parameter. The SSCRB is the proper bound to be calculated to assess the efficiency of any estimator in such a scenario. Numerical results have shown that the best performance are achieved by exploiting the MUSIC algorithm together with the Tyler's or Huber's $M$-estimate of the scatter matrix. However, none of the considered estimators is an efficient one w.r.t. the SSCRB. This open problem, together with the experimental validation of the measurement model adopted in this paper, will be addressed in future works.

%\subsection{Figures and Tables}
%\paragraph{Positioning Figures and Tables} Place figures and tables at the top and 
%bottom of columns. Avoid placing them in the middle of columns. Large 
%figures and tables may span across both columns. Figure captions should be 
%below the figures; table heads should appear above the tables. Insert 
%figures and tables after they are cited in the text. Use the abbreviation 
%``Fig.~\ref{fig}'', even at the beginning of a sentence.
%
%\begin{table}[htbp]
%\caption{Table Type Styles}
%\begin{center}
%\begin{tabular}{|c|c|c|c|}
%\hline
%\textbf{Table}&\multicolumn{3}{|c|}{\textbf{Table Column Head}} \\
%\cline{2-4} 
%\textbf{Head} & \textbf{\textit{Table column subhead}}& \textbf{\textit{Subhead}}& \textbf{\textit{Subhead}} \\
%\hline
%copy& More table copy$^{\mathrm{a}}$& &  \\
%\hline
%\multicolumn{4}{l}{$^{\mathrm{a}}$Sample of a Table footnote.}
%\end{tabular}
%\label{tab1}
%\end{center}
%\end{table}

%\begin{figure}[htbp]
%\centerline{\includegraphics{fig1.png}}
%\caption{Example of a figure caption.}
%\label{fig}
%\end{figure}

\bibliographystyle{IEEEtran}
\bibliography{ref_EUSIPCO_2019}

\end{document}